\documentclass[jgr]{agutex}
\usepackage[dvips]{graphicx}
\usepackage{amsfonts}
\usepackage{amssymb}
\usepackage{verbatim}

\authorrunninghead{RUSSELL ET AL.}

\titlerunninghead{PRODUCTION OF SMALL-SCALE ALFV\'EN WAVES}

\authoraddr{A. J. B. Russell,
SUPA School of Physics and Astronomy, Kelvin Building, University of Glasgow, Glasgow, G12 8QQ, Scotland, UK.
(alexander.russell@glasgow.ac.uk)}
\authoraddr{A. N. Wright, 
School of Mathematics and Statistics, University of St Andrews, North Haugh, St Andrews, Fife, KY16 9SS, Scotland, UK.
(andy@mcs.st-andrews.ac.uk)}
\authoraddr{A. V. Streltsov, 
Department of Physical Sciences, Embry-Riddle Aeronautical University, 600 S. Clyde Morris Blvd., Daytona Beach, FL 32114, USA.
(streltsa@erau.edu)}

\begin{document}

\title{Production of small-scale Alfv\'en waves by ionospheric depletion, nonlinear magnetosphere-ionosphere coupling and phase mixing}
\authors{A. J. B. Russell, \altaffilmark{1}
A. N. Wright, \altaffilmark{2}
and A. V. Streltsov\altaffilmark{3}}
\altaffiltext{1}{School of Physics and Astronomy, Kelvin Building, University of Glasgow, Glasgow, G12 8QQ, Scotland, UK.}
\altaffiltext{2}{School of Mathematics and Statistics, University of St Andrews, North Haugh, St Andrews, Fife, KY16 9SS, Scotland, UK.}
\altaffiltext{3}{Department of Physical Sciences, Embry-Riddle Aeronautical University, 600 S. Clyde Morris Blvd., Daytona Beach, FL 32114, USA.}
\date{Received <date> / Accepted <date>}

\begin{abstract}
Rockets and satellites have previously 
observed small-scale Alfv\'en waves inside large-scale downward field-aligned currents
and numerical simulations have associated their formation with self-consistent magnetosphere-ionosphere coupling. 
The origin of these waves was previously attributed to ionospheric feedback instability, 
however we show that they arise in numerical experiments in which the instability is excluded.
A new interpretation is proposed in which strong ionospheric depletion and associated current broadening 
(a nonlinear steepening/wavebreaking process) form magnetosphere-ionosphere waves inside a downward current region
and these oscillations drive upgoing inertial Alfv\'en waves in the overlying plasma.
The resulting waves are governed by characteristic periods, 
which are a good match to previously observed periods for reasonable assumed conditions.
Meanwhile, wavelengths perpendicular to the magnetic field initially map to an ionospheric scale
comparable to the electron inertial length for the low-altitude magnetosphere,
but become shorter with time due to frequency-based phase mixing of boundary waves (a new manifestation of phase mixing).
Under suitable conditions, these could act as seeds for the ionospheric feedback instability.
\end{abstract}

\begin{article}

\section{Introduction}\label{sec:intro}
Small-scale Alfv\'en waves are regularly observed inside large-scale magnetospheric current systems 
\citep[e.g. review by][and references therein]{2000Stasiewicz}
and several studies have shown an association with downward current regions 
\citep[e.g.]{2004Karlsson,2005Keiling,2004Johansson,2008Wright}.
They are speculated to play a role in electron acceleration \citep{2002Chaston} and in modifying F-region densities 
\citep{1990Boehm,2008StreltsovLotko};
however, their origin remains a subject of discussion.

A particularly curious characteristic of such waves is their period, 
which was estimated for one event as 20 to 40 seconds \citep{2004Karlsson}.
It is difficult to associate such periods with ionospheric timescales,
such as trapping in the ionospheric Alfv\'en resonator (IAR), because these tend to be about one second or less.
Furthermore, waves are not confined to the IAR, being observed at altitudes up to 4 ${\rm R}_{\rm E}$.
Attempts to associate waves with magnetospheric trapping likewise encounter difficulties, 
because typical magnetospheric travel times are hundreds of seconds (e.g. ultra-low frequency waves in the Pc5 range)
and waves occur near the boundary between open and closed field-lines.
Thus attempts to give these waves a purely magnetospheric or purely ionospheric explanation are unsuccessful.

One line of inquiry that has been promising is the premise that small-scale waves might be produced 
by self-consistent magnetosphere-ionosphere (M-I) coupling.
This was investigated by \citet{2004StreltsovLotko,2008StreltsovKarlsson}, 
whose numerical experiments demonstrated that waves resembling observations are produced by a simulated system's response 
to large-scale field-aligned currents (FACs) via the ionospheric feedback mechanism (IFM).

In their original paper, \citet{2004StreltsovLotko} identified instability
as the most likely explanation for production of small-scale waves.
This explanation is appealing because waves first appeared inside a region where the underlying E-region 
was depleted (low Pedersen conductance) and electric field was strong, and these conditions are known to favor growth of 
ionospheric feedback instability (IFI) \citep{1970Atkinson,1973HolzerSato,1978Sato,1991Lysak,2002LysakSong}.
It was therefore proposed that the experiment might develop small-scale waves through amplification of initial seed perturbations by IFI.

Some features of the waves observed in numerical experiments and the real magnetosphere do not fit well with an IFI explanation.
Feedback instability requires a trapping region so that waves may be overreflected from an underlying E-region many times.
Trapping should therefore be expected to leave a signature on waves that have grown through IFI, 
e.g. by imprinting a period or magnetic-field aligned length scale on the resulting waves.
The waves produced in the experiments by \citet{2004StreltsovLotko} did not show
such periodic structure along the field-aligned direction,
and observed wave periods are an order of magnitude longer than typical IAR periods.

The evolution of small-scale waves in the magnetosphere and ionosphere and their impact on auroral processes has also been studied
by, e.g., \citet{1990Seyler,2002Chaston,2004Genot,2008LysakSong} and references therein, often using highly sophisticated models.
These studies, however, all imposed either some short timescale or wavelength on 
the system through their driver or initial conditions.
Therefore, although they offer valuable insights into the evolution and impacts of small-scale Alfv\'en waves,
they did not address how the waves are initially produced and how the system imposes timescales and wavelengths,
which is the focus of this work.

This paper aims to establish an alternative explanation for the origin of
small-scale Alfv\'en waves produced by nonlinear M-I coupling, 
and to show how wave properties are determined.
First, we demonstrate that feedback instability in not necessary to produce small-scale waves in downward current regions,
by presenting a simplified numerical experiment which excludes feedback instability while retaining the feedback mechanism.
We then offer a new interpretation in terms of ionospheric depletion, 
wavebreaking (nonlinear steepening), magnetosphere-ionosphere waves and frequency-based phase mixing.
This novel explanation draws on recent theoretical advances in the topic of magnetosphere-ionosphere coupling,
which are linked for the first time.
Using these, we are able to clarify the process by which M-I coupling can produce small-scale waves, 
and highlight formulas for wave periods and wavelengths.

\section{Modeling}\label{sec:model}
The first goal of this paper is to test the hypothesis that small-scale Alfv\'en waves seen in numerical M-I coupling experiments
are produced by the action of ionospheric feedback instability.

IFI has two essential ingredients.  
First of all, the ionospheric feedback mechanism is required to act in the presence of an ambient electric field, 
which may be due to the presence of a large-scale current system.
Under these conditions, small-scale Alfv\'en waves, incident on the E-region from above,
can be overreflected if they have suitable properties \citep{1984TrakhtengertsFeldsein,2002LysakSong,2012RussellWright}.
In these cases, the reflected small-scale wave has greater amplitude than the incident wave,
increased energy flux being accounted for by a reduction in the ionospheric heating caused by the ambient electric field.
Instability can follow if a second ingredient is present: reflection of upgoing Alfv\'en waves at some location above the E-region,
e.g. from the conjugate ionosphere
or from gradients in Alfv\'en speed below the peak at about $1\mbox{ R}_E$
in which case waves are partially trapped inside the IAR.
If this second reflection produces a downgoing wave whose amplitude is greater than the initial incident wave,
then successive cycles of overreflection at the E-region and partial reflection at an overlying altitude cause
waves to grow exponentially and a system seeded with low-amplitude perturbations becomes dominated by the fastest growing mode
\citep{1970Atkinson,1973HolzerSato,1978Sato,1984TrakhtengertsFeldsein,1991Lysak,2002LysakSong}.

Since both IFM and wave trapping are required for IFI, 
the hypothesis that IFI is responsible for producing the small-scale waves seen by \citet{2004StreltsovLotko}
can be tested using a simplified model that includes the essential elements of IFM, 
but which is designed to avoid reflections above the E-region, thus preventing IFI.

\begin{figure}
 \resizebox{0.8\hsize}{!}{\includegraphics{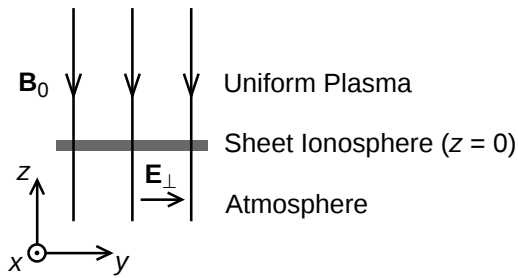}}
 \caption{Geometry of simplified electrodynamic magnetosphere-ionosphere coupling model.
          }\label{fig:model}
\end{figure}

To this end, we have investigated what happens when a sheet ionosphere is coupled to a uniform overlying plasma
and the system driven with large-scale currents in the form of a large-scale incident Alfv\'en wave.
The sheet ionosphere description is mathematically valid provided the electric field skin depth in the ionosphere is 
larger than or comparable to the ionospheric thickness, generally true for frequencies below 100 Hz \citep{1991Lysak}.
Matters are simplified by using a 2D model, 
sketched in Figure \ref{fig:model}, in which $z$ is the vertical coordinate, 
$x$ is an invariant horizontal direction, and $y$ completes the Cartesian system.
Equilibrium magnetic field is assumed vertical (appropriate for polar latitudes and low altitudes) 
and the direction of the horizontal electric field defines the $y$-direction.
We are interested in timescales much longer than cyclotron or plasma periods, so the plasma is described as a single fluid.
Wave amplitudes are assumed small enough for nonlinear effects to be unimportant in the magnetospheric plasma region,
because these are unlikely to play a role in producing small scales via IFM;
the majority of current closure by Pedersen currents is assumed to occur in the sheet ionosphere,
so that Pedersen conductivity can be neglected in the overlying plasma;
and we neglect dissipation due to Landau damping, viscosity and resistivity, 
which may affect the long term evolution of small-scale Alfv\'en waves but are unlikely to be important for their creation.
The overlying plasma region is therefore described using a single fluid model 
in which dynamics can be described in terms of inertial Alfv\'en waves.

IFM is included by solving the ionospheric response to field aligned current through the height-integrated continuity equation:
\begin{eqnarray}
\frac{\partial N}{\partial t}&=&\frac{j_{z}}{e}+\frac{\alpha}{h}\left(N_e^2-N^2\right),\label{eq:dNdt}
\end{eqnarray}
where $N$ is the height-integrated ionospheric plasma density, 
$j_{z}$ is the field-aligned current at the top of the sheet ionosphere,
$e$ is the fundamental charge,
$\alpha$ is an effective recombination coefficient \citep{1997Brekke},
and $h$ is the ionospheric thickness.
The source term $\alpha N_e^2/h$ represents background ionization, 
the loss term $-\alpha N^2/h$ represents recombination
and $j_{z}/e$ accounts for addition (removal) of electrons by upward (downward) FACs.

Ohmic heating can change recombination rates, and this can be important for F-region modeling, 
where the low densities allow for significant changes on timescales as short as a few minutes 
\citep{1978StMauriceTorr,2010Zettergren,2012ZettergrenSemeter}.
In the E-region, however, the much greater thermal mass means that 
$\alpha$ may be treated as constant for the timescales we study (several to at most tens of minutes).
Heating of the F-region may affect the Alfv\'en speed there, and we discuss the consequences of this in our discussion, 
but our primary conclusions regarding creation of small-scale waves and their properties are unaffected.

The feedback loop of IFM is completed by imposing a boundary condition at the base of the overlying plasma that depends on ionospheric number density.
This is derived using current closure and takes the form,
\begin{eqnarray}
 b_x=(\mu_0 e M_P)NE_y,\label{eq:IBC}
\end{eqnarray}  
for our 2D model, where $M_P$ is the Pedersen mobility  
defined by $\Sigma_P=eM_PN$ with $\Sigma_P$ the effective height-integrated Pedersen conductance.
$M_P$ is taken constant in our study.

Magnetospheric equations (describing Alfv\'en waves in a fully ionized plasma with electron inertia, under the assumptions already stated) 
are as follows:
\begin{eqnarray}
 \frac{\partial b_x}{\partial t}&=&\frac{\partial E_y}{\partial z}-\frac{\partial E_z}{\partial y},\label{eq:dbx}\\
 \frac{\partial E_y}{\partial t}&=&v_A^2\frac{\partial b_x}{\partial z},\label{eq:dEy}\\
 \frac{\partial E_z}{\partial t}&=&-c^2\left(\mu_0j_z+\frac{\partial b_x}{\partial y}\right),\label{eq:dEz}\\
 \frac{\partial j_z}{\partial t}&=&\frac{E_z}{\mu_0\lambda_e^2}.\label{eq:djz}
\end{eqnarray}
A derivation of these equations is given by \citet{2010Russell} and
they can also be found as a limiting case of the equations used by \citet{2008LysakSong}.

The whole system is solved numerically using an explicit leapfrog trapezoidal scheme
and centered finite differences \citep{2010Russell}.
In equation (\ref{eq:dEz}), the parameter $c$ is assigned an artificial value that ensures the 
electron plasma frequency, $\omega_{pe}=c/\lambda_e$, 
is much greater than the angular frequency of the oscillations of interest 
(ensuring proper inertial Alfv\'en wave behavior)
but can also be resolved with a reasonable timestep (ensuring numerical stability).
This technique has been previously used by \citet{2008LysakSong} and is discussed by \citet{2010Russell}.
Reflections from the upper boundary (which could potentially lead to a numerical form of IFI) are avoided
by positioning it at a distance that ensures waves cannot leave the
subsection of the simulation domain shown in our figures, reflect from the upper boundary and return to the subdomain
within the simulation runtime. Thus, only part of the magnetospheric domain is shown in our figures.

The resulting model is much simpler than others that have been used to describe the ionosphere-thermosphere system,
e.g. \citet{1990Seyler,1997Dreher,2001Zhu,2004StreltsovLotko,2008Sydorenko,2011Chaston,2012ZettergrenSemeter},
however, for the purposes of this paper it offers some important advantages.
First, it ensures that any small-scale waves produced are created by IFM but not IFI, providing a rigorous test of whether or not IFI is required,
and second, it allows the production of small-scale waves to be studied in isolation from their later development,
which might otherwise obscure the details of their origin.
Simple models have a strong track record for showcasing the essential aspects of a problem,
and, if designed appropriately, they are valuable for developing new interpretations and identifying formulas 
that remain a useful guide when additional physics is later added.
The task of adding greater complexity, and seeing how these fundamental behaviors are altered, is left for the future.

It is at times useful to refer to simulations that do not include electron inertia in the magnetospheric domain,
particularly to clarify how small scales are produced in the absence of IFI.
These are performed using an approach proposed by \citet{2007CranMcGreehin} which reduces the model to a single governing equation
that is solved numerically using a leapfrog trapezoidal scheme with one-sided finite differences \citep{2010RussellWright}.
Apart from the neglect of electron inertia, 
the assumptions for this ideal MHD model are the same as for the electron-inertial single-fluid model.

\section{Waves Produced in Absence of Instability}\label{sec:waves_without_ifi}
  
To determine whether small-scale waves are produced by IFM in the absence of IFI, 
the electron-inertial M-I model described in Section \ref{sec:model} was driven with a system of large-scale currents and allowed to evolve.
The currents were created by specifying a large-scale incident Alfv\'en wave
that shears the background magnetic field to produce one channel of upward FAC and one channel of downward FAC.
(This form of driving is equivalent to that used by \citet{2004StreltsovLotko}.)
The incident wave does not contain small scales and remains constant after a short ramping transient.
Current systems like this may be produced by processes in the magnetosphere, far from the ionosphere, but their origin is not considered here.
Suffice to say that currents do form, and we examine the M-I system response.

\begin{figure}
 \resizebox{0.8\hsize}{!}{\includegraphics{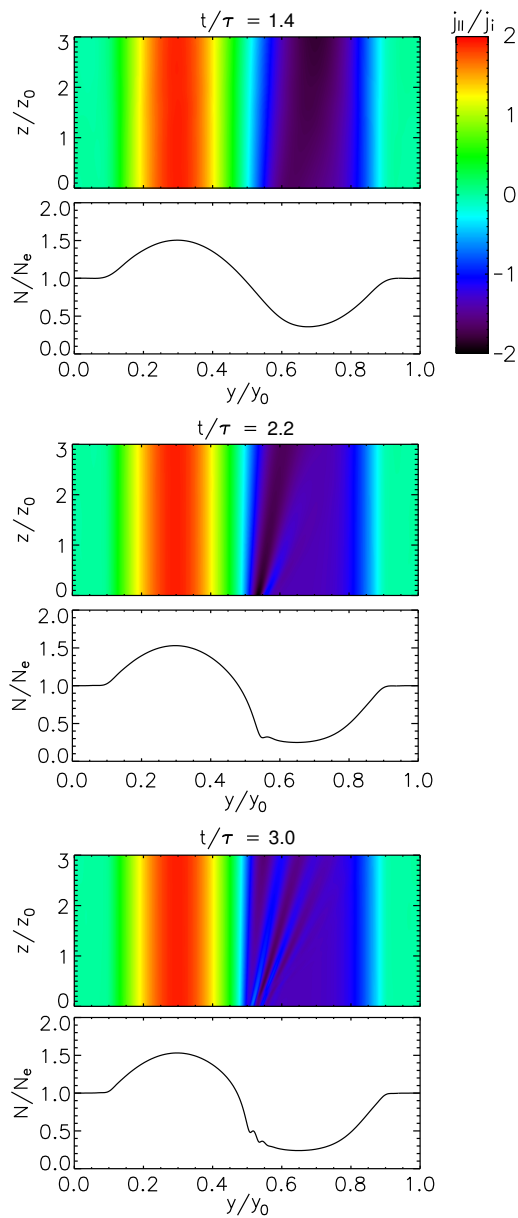}}
 \caption{Response of a uniform plasma with finite electron inertia and a sheet ionosphere to a large scale current.  
         Color shows field-aligned current, $j_{z}$, and line plots show height-integrated ionospheric plasma density, $N$.  
          }\label{fig:inertial}
\end{figure}

The evolution of our simulation is shown by Figure \ref{fig:inertial}, which presents three snapshots of field-aligned current density 
(normalized to the maximum current density of the incident driving wave, $j_i$)
and height-integrated ionospheric number density (normalized to the equilibrium value in the absence of field-aligned currents, $N_e$).
The sheet ionosphere is positioned at $z=0$, which corresponds to an altitude of approximately 110 km.
Times are normalized by a depletion timescale, $\tau=eN_e/j_i$,
this choice lending itself to the present study better than 
the advection timescale used for normalization by \citet{2012RussellWright}.
Horizontal distances are normalized by the width of the simulation domain, $y_0$, 
and vertical distances are normalized by $z_0=v_A\tau$ where $v_A$ is the Alfv\'en speed.  Typically, $z_0 \gg y_0$.

Examining Figure \ref{fig:inertial},
at $t/\tau=1.4$, the large-scale current system is well established and ionospheric number density is evolving in response to the currents.
One can see that $N$ has been increased in the upward current channel (centered on $y/y_0=0.3$), as electrons are deposited there, 
and decreased in the downward current channel (centered on $y/y_0=0.7$), as electrons are removed there.

At $t/\tau=2.2$ the upward current channel has very nearly reached a new steady state
(the maximum value of $N$ is within 0.3\% of its steady-state and converging slowly to it), however, 
the downward current channel and associated ionospheric depletion region are in the process of broadening 
on the left-hand side, where they expand in the direction of the horizontal electric field.
As they widen, $N$ becomes steep and a ripple begins to form around $y/y_0=0.55$,
at this time looking like a small tooth at the left-hand edge of the depleted region.

At later times, e.g. $t/\tau=3.0$, small-scale waves are clearly present.
Small-scale changes in $N$, apparent between $y/y_0=0.48$ and $y/y_0=0.58$, 
correspond to small-scale changes in ionospheric reflectivity, 
so when the large-scale incident Alfv\'en wave driving the system reflects from the ionosphere,
small-scale upgoing Alfv\'en waves are formed 
(visible in the color plot of $j_{z}$ at $t/\tau=3.0$ 
as alternating bands of blue and purple in the downward current channel with short horizontal wavelength).
These are inertial Alfv\'en waves and they carry small scales that originate at $z=0$ into the magnetosphere.
Their phase-speed, transverse to the magnetic field, is in the direction of the large-scale electric field (to the left) 
and their group velocity, transverse to the magnetic field, is oppositely directed (to the right),
causing them to spread out over the downward current channel with increasing altitude.

These features are similar to those seen at early times in the experiments of \citet{2004StreltsovLotko}.
In particular, boundary waves form just inside the downward current channel, at the edge adjacent to the upward current channel,
and the same tooth-like feature is seen in both experiments (see their Figures 3 and 4).
We therefore conclude that small-scale waves are formed by the same process in both experiments,
even though the experiments of \citet{2004StreltsovLotko} included many more aspects of the M-I system.
This reassures us that although the present model is simpler, it does include the essential elements we wish to understand.

These results demonstrate that small-scale waves can be produced by nonlinear M-I coupling in the absence of instability,
clarifying the mechanism of their origin.
In particular, they do not arise through amplification of noise by IFI as suggested previously.
We now outline a new interpretation that proposes how small-scale waves may be formed and what their properties will be.

\section{Small Scales from Ionospheric Depletion}\label{sec:ideal}
Formation of small scales appears to involve ionospheric depletion 
and horizontal expansion of the depleted region and associated downward current channel.
It is therefore worthwhile to examine the conditions under which broadening occurs and
the manner in which it happens, with particular focus on the generation of small scales.

The theory of ionospheric depletion and cavity/current broadening has previously been studied by 
\citet{1995Doe,1996BlixtBrekke,1998KarlssonMarklund,2007CranMcGreehin,2010RussellWright,2012ZettergrenSemeter}
and we quote their work for the condition for broadening.
These authors have shown that the coupled M-I system has two distinct types of evolution, 
depending on the maximum strength of downward current density.
Examining equation (\ref{eq:dNdt}), background ionization can balance downward FAC provided that 
$|j_{z}|<j_c$, where
\begin{eqnarray}
 j_c&=&\frac{\alpha N_e^2 e}{h}.\label{eq:jc}
\end{eqnarray}
Consequently, the ionospheric response to an incident Alfv\'en wave depends on the strength of the current density the 
large-scale incident wave creates in relation to this threshold.  

For an ionosphere that is initially highly conducting, the maximum current density at its top is approximately $2j_i$,
where $j_i={\rm max}({\rm abs}( \partial (\delta B_i)/\partial y))/\mu_0$ 
is the maximum parallel current density of the incident Alfv\'en wave.
If $|2j_i|<j_{c}$, then the coupled system evolves smoothly to a new partially-depleted steady state \citep{2007CranMcGreehin,2010RussellWright}.

\begin{figure}
 \resizebox{0.8\hsize}{!}{\includegraphics{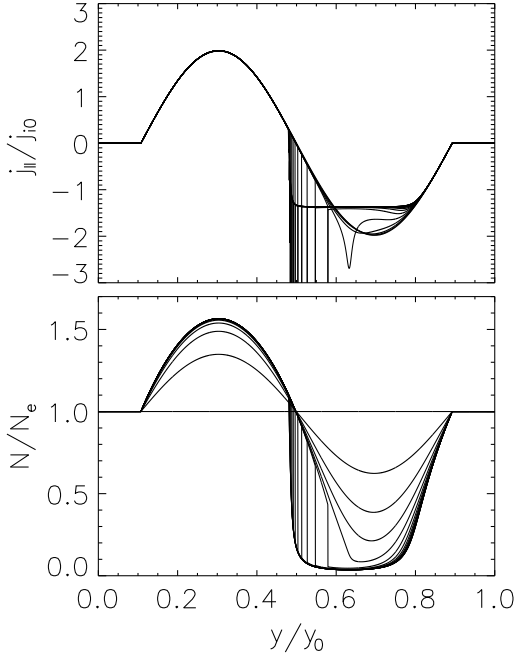}}
 \caption{Evolution with incident downward current satisfying $|2j_i|>j_{c}$ and no electron inertia, 
          showing formation of a discontinuity in ionospheric number density and a corresponding current sheet.
          Snapshots show ionospheric plasma density (bottom) and field-aligned current at the top of the ionosphere (top).
          }\label{fig:ideal}
\end{figure}

If $|2j_i|>j_{c}$, then dynamics are more dramatic.
Figure \ref{fig:ideal} shows a simulated system response to a downward current density that exceeds 
the critical value and is therefore unsustainable.
Starting with an initially uniform ionosphere,
downward FAC strongly depletes ionospheric plasma to such an extent 
that ionospheric reflectivity is significantly reduced,
decreasing downward current density;
hence, the weakened downward FAC is forced to broaden to close the upward FAC.
Here, electron inertia is neglected and as the downward FAC broadens, a moving density discontinuity forms in the ionosphere, 
corresponding to a moving current-sheet in the magnetosphere.

The ideal simulation presented in Figure \ref{fig:ideal} uses an initial condition with $\Sigma_P/\Sigma_A=100$, 
for which the features of broadening are very clear. 
Our inertial model is not capable of solving under these conditions
because deep depletion associated with high values of $\Sigma_P/\Sigma_A$ can trigger numerical instability through our boundary condition.
Consequently, the simulation shown in Figure \ref{fig:inertial} uses $\Sigma_P/\Sigma_A=10$ for stability,
while the simulation shown in Figure \ref{fig:inertial} uses a higher value for improved clarity.
This means that these figures should not be compared quantitatively,
although a qualitative comparison is very valuable.

The aspects of broadening outlined so far have been described previously by \citet{2007CranMcGreehin}, 
however recent theoretical developments allow us to add some new comments about the broadening process.
First, it is interesting to consider broadening in the context of recent knowledge
that the governing equations for IFM reduce to an advection equation in the absence of electron inertia \citep{2012RussellWright}.
Thus, dynamics are governed by a characteristic advection speed, which, for a 1D sheet ionosphere, 
produces motion in the direction of the electric field at a speed
\begin{eqnarray}
 v_{MI}&=&\frac{M_PE_y}{1+\Sigma_P/\Sigma_A}\label{eq:v_MI}
\end{eqnarray}
\citep{2012RussellWright}.
Preexisting arguments that explain broadening only by need to provide current closure (such as that outlined above) 
do not lead one to expect asymmetry.
However, since the advective properties only permit motion in the direction of the electric field,
it follows that broadening can only happen at an edge where the electric field points away from the depleted region.
This explains why the depleted region and downward current channel may broaden only on one side.

The discontinuity itself moves at the M-I advection speed derived for discontinuities by \citet{2012RussellWright},
which is the geometric mean of the values of $v_{MI}$ to either side of the discontinuity.
This gives the rate of expansion and is applied here to ionospheric depletion for the first time.

\begin{figure}
 \resizebox{\hsize}{!}{\includegraphics{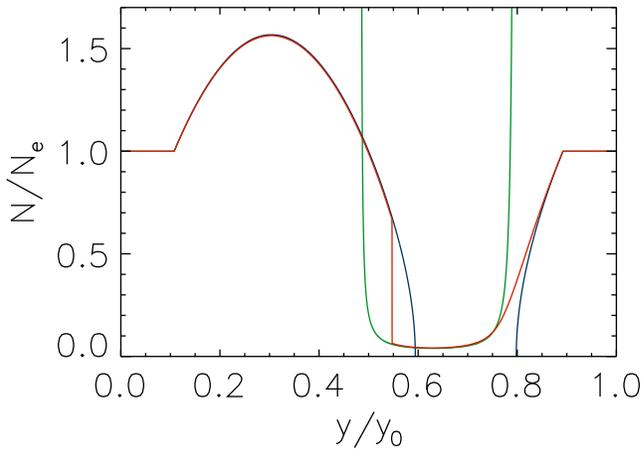}}
 \caption{Snapshot of density discontinuity formed in absence of electron inertia (red),
          upper steady state (blue) and lower steady state (green).
          }\label{fig:discontinuity}
\end{figure}

It is also informative to compare a single snapshot from Figure \ref{fig:ideal} 
to the self-consistent M-I steady states obtained by \citet{2010RussellWright}.
Such a comparison in provided in Figure \ref{fig:discontinuity}.
From this, one sees that the discontinuity steps between an ``upper" steady state 
(computed under the assumption $\Sigma_P \gg \Sigma_A$ and plotted in blue)
and a ``lower" steady state (computed under the assumption $N^2 \ll N_e^2$ and plotted in green).
When the discontinuity first forms, it does so where the upper and lower steady states cross 
inside the downward current channel (at $y/y_0=0.59$).
Thereafter, the discontinuity sweeps to the left across the region 
where both steady states are valid and stops where they cross again (at $y/y_0=0.49$).
Thus, these solutions indicate the early and final widths of the depleted region, 
as well as the height of the discontinuity at any position.

\begin{figure}
 \resizebox{0.8\hsize}{!}{\includegraphics{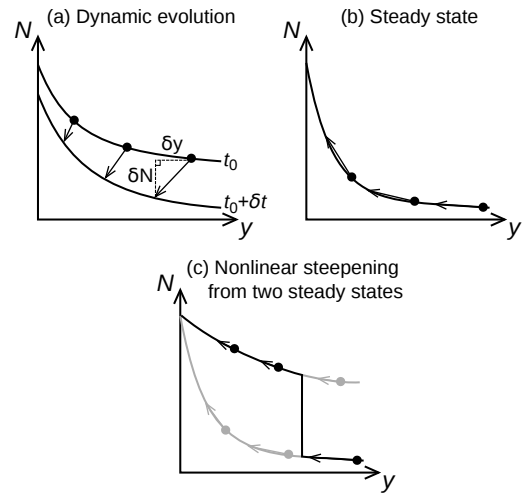}}
 \caption{Trajectories of points in $(y,N)$ space under advection.
          (a) Motion of points during dynamic evolution, giving $\partial N / \partial t \neq 0$.
          (b) Motion of points along a steady state curve, maintaining $\partial N / \partial t = 0$.
          (c) Where two steady states exist, points move more rapidly along the lower steady state,
              so a moving discontinuity forms to preserve a single valued solution.
          }\label{fig:trajectories}
\end{figure}

The combination of advective properties with the existence of two steady state solutions, 
leads us to propose that broadening is essentially an M-I wavebreaking (nonlinear steepening) process.
A useful visualization of advection, sketched in Figure \ref{fig:trajectories},
is to imagine the motion of points over time in $(y,N)$ space
under the action of advection in $y$ ($dy/dt=v_{MI}$) and changes in $N$ due to the convective derivative, $dN/dt$.
In an infinitesimal time $\delta t$, a point initially at $(y,N)$ moves to $(y+\delta y, N+\delta N)$, 
with $\delta y = v_{MI}(y,N)\delta t$ and $\delta N = (dN/dt)\delta t$.
Thus, in dynamic evolution (panel (a) of Figure \ref{fig:trajectories}), 
$N(y,t)$ evolves as the curve is carried by the motion of the points it threads.
In this context, a steady state solution (panel (b) of Figure \ref{fig:trajectories})
is a trajectory common to all points lying on it,
hence, although points move along the steady state, the curve threading them does not change, giving $\partial N/\partial t=0$.

Referring to Figure \ref{fig:ideal}, our simulation commences with dynamic evolution, 
as points starting on the initial condition move towards the nearest steady state.
In the upward current channel and the at edges of downward current channel, this is the upper steady state.
In the center of the downward current channel, it is the lower steady state.
At the right hand edge of downward channel, the upper and lower steady states break down at a common location
and the true steady state is a matching between these which 
can be obtained using the asymptotic method of \citet{2010RussellWright}.
At the left hand edge of downward channel, the upper steady state is encountered first,
so points in $(y,N)$ space pause here although there is also a valid lower steady state at lower $N$.

Once points have moved on to a steady state curve, they move along it, with motion in $y$ determined by the advection speed.
The vital property responsible for wavebreaking is this: 
examining equation \ref{eq:v_MI}, a point at $(y,N)$ 
moves more rapidly than a point at the same $y$ but greater $N$
(for which $\Sigma_P/\Sigma_A$ is higher).
Thus, in the region where there are two valid steady states, 
points following the upper steady state move more slowly than, and are overtaken by, points moving along the lower steady state
(panel (c) of Figure \ref{fig:trajectories}).
When points catch up in $y$, a discontinuity forms to preserve a single valued physical solution.
This wavebreaking (nonlinear steepening) picture presents a new conceptual view of the 
broadening that occurs in response to ionospheric depletion, 
connecting broadening with the advective properties of M-I coupling 
and the existence of two steady state solutions.

We conclude from this section that the coupled M-I system produces small horizontal scales  
in response to strong downward FAC that exceeds the critical threshold $j_{c}$,
and that it does this through ionospheric depletion and broadening of downward FAC,
which can be considered as a wavebreaking (nonlinear steepening) process,
that occurs because of the existence of two steady state solutions.

\section{Wave Properties and Evolution}\label{sec:waves}
In reality, gradient length scales do not collapse to zero because small-scale physics prevents this.
For the present M-I coupling work, the appropriate modification is to include electron inertia in the magnetosphere.

The changes due to electron inertia can be seen by comparing the qualitative features of Figure \ref{fig:inertial} 
(finite electron inertial length, $\lambda_e/y_0=0.025$) with those of Figure \ref{fig:ideal} ($\lambda_e=0$). 
At early times, length scales seen in Figure \ref{fig:inertial} are much greater than $\lambda_e$, 
so the simulation follows the ideal MHD evolution described in Section \ref{sec:ideal}.
When wavebreaking occurs, a steep transition appears and moves to the left, 
but it has finite width comparable to $\lambda_e$ instead of being a discontinuity.
This broadening front is trailed by an undershoot, which later develops into a train of undershoots and overshoots.
Time evolution at a fixed location is gradual until the broadening front passes, 
its movement causing a rapid decrease in $N$ that produces oscillations.
Such effects are to be expected by analogy to MHD shocks, 
which exhibit very similar features when the dominant small-scale physics is dispersive.

The wake of undershoots and overshoots that trails the broadening front in $N$ is of great interest 
because these disturbances drive the upgoing inertial Alfv\'en waves whose origin and properties we aim to explain
When the incident large-scale wave driving the system reflects from an ionosphere with 
small-scale variations in ionospheric reflectivity (due to small scale variations in $N$), 
these variations impose their scale and period on the reflected wave,
which carries these properties into the magnetosphere.

Wave-like behaviors resulting from self-consistent electrodynamic M-I coupling
were recently described by \citet{2012RussellWright}, 
who suggested the name ``magnetosphere-ionosphere (M-I) waves" for the phenomena
and also identified nonlinear steepening as a means of accessing small scales.
The properties of M-I waves are shared by the small-scale disturbances produced on the boundary in the present study,
and it is useful to refer to them for interpretation.
The current work does, however, advance our understanding of these waves in three important ways:
(i) it demonstrates that small-scale M-I waves can be be produced from a simple large-scale driver
by the depletion mechanism outlined above;
(ii) the spatially varying steady state introduces M-I wave phase mixing (discussed below)
that reduces length scales further and was not considered by \citet{2012RussellWright};
and (iii) we introduce a simple formula for the largest inertial wavelength produced by wavebreaking.
We also note that \citet{2012RussellWright} did not apply their results to the generation of small-scale Alfv\'en waves
launched into the magnetosphere following ionospheric depletion, which is the focus of this paper.

According to the theory of M-I waves,
as electron inertial becomes significant, the advection that occurs for $\lambda_e=0$ is altered,
giving way to oscillation at a characteristic frequency for wavelengths comparable to or less than $\lambda_e$
(see Figure 4 of \citet{2012RussellWright}).
This frequency is given by 
\begin{eqnarray}
 f_{MI} &=& \frac{M_PE_y}{2\pi\lambda_e\left(\Sigma_{P}/\Sigma_{A}\right)}.\label{eq:frequency}
\end{eqnarray}
Meanwhile, wave amplitudes decay on the ionospheric recombination time,
\begin{eqnarray}
 \tau_{decay}=\frac{h}{2\alpha N}.\label{eq:decay}
\end{eqnarray}

\begin{figure}
 \resizebox{0.8\hsize}{!}{\includegraphics{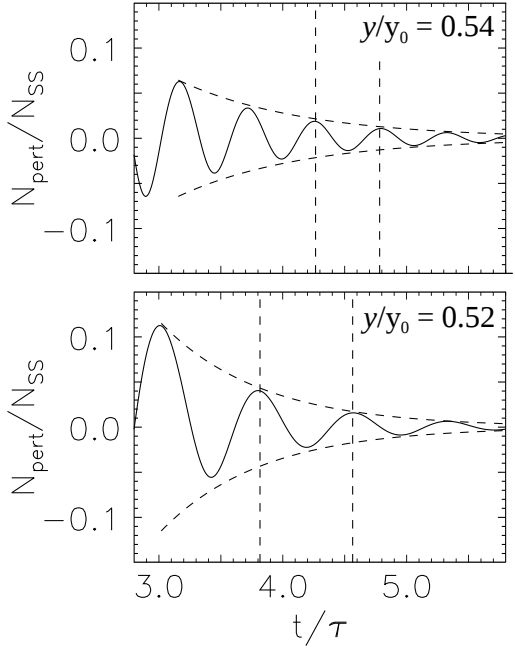}}
 \caption{Perturbation in height-integrated ionospheric number density relative to steady state value, 
          at $y/y_0=0.54$ (top) and $y/y_0=0.52$ (bottom).
          For each panel, vertical dashed lines are separated by $\tau_{MI}/\tau$, 
          and the dashed envelope is calculated for exponential decay at the recombinative e-folding time $\tau_{decay}/\tau$,
          where these times are computed from steady state values at each location.
          }\label{fig:periods}
\end{figure}

Figure \ref{fig:periods} plots the relative perturbations of $N$ at $y/y_0=0.52$ and $y/y_0=0.54$ 
between times $t/\tau=2.8$ and $t/\tau=5.8$.  The time interval is chosen to focus 
on the evolution of the oscillations and not the gradual evolution and rapid fall in $N$ that precedes them.
Vertical dashed lines are separated by $\tau_{MI}/\tau$ where $\tau_{MI}=1/f_{MI}$ is the period expected for M-I waves,
and the amplitude envelope corresponds to exponential decay with an e-folding time of $\tau_{decay}$.
Both $f_{MI}$ and $\tau_{decay}$ are evaluated using steady state values of $E_y$ and $N$ about which the waves oscillate.
The formulas for $f_{MI}$ and $\tau_{decay}$ (equations (\ref{eq:frequency}) and (\ref{eq:decay}) respectively)
are an excellent match to the oscillations, 
even though background quantities vary in $y$ and the formulas were derived using normal mode analysis,
effectively looking for local linear modes about a global equilibrium.

The high level of agreement gives us confidence in connecting the properties of inertial M-I waves 
with small-scale waves produced in response to strong downward current.
This link, which is new, is significant because: 
(i) it provides formulas for the properties of small-scale waves produced by the mechanism described in Section \ref{sec:ideal};
and (ii) it identifies a natural process capable of creating M-I waves, which furthers their study.
Equation (\ref{eq:frequency}) is easily used and gives frequencies that are likely to dominate a low quality IAR,
(i.e. an IAR for which Alfv\'en speed gradients do not produce substantial reflectivity
for the wave frequencies under consideration, implying long growth times for IFI
and allowing substantial transmission of upgoing waves to the magnetosphere).
A high quality IAR with large IFI growth rate could also exhibit the conventional fastest growing mode frequency,
computation of which requires more cumbersome numerical simulations \citep{Streltsov_substorm}.

\begin{figure}
 \resizebox{\hsize}{!}{\includegraphics{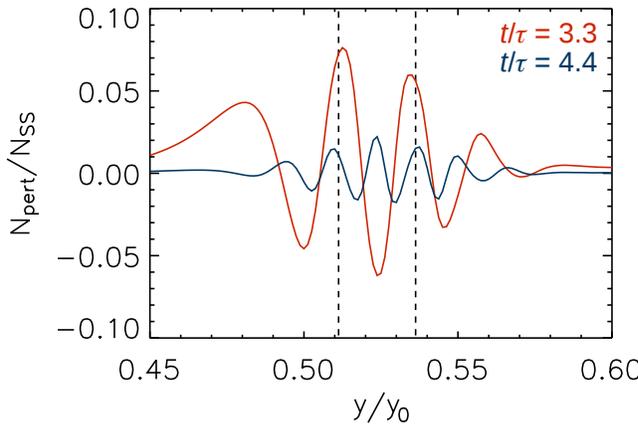}}
 \caption{Perturbation in height-integrated ionospheric number density relative to steady state value,
          at $t/\tau=3.3$ (red) and $t/\tau=4.4$ (blue) with
          dashed lines showing the electron inertial length.
          }\label{fig:phase-mixing}
\end{figure}

It remains to determine the wavelength of the M-I waves on the boundary.
Figure \ref{fig:phase-mixing} shows a close-up of the perturbation in $N$ at two different times,
with the electron inertial length indicated.
The waves clearly have a wavelength (measured along the boundary) comparable to $\lambda_e$ and their group velocity is negligible.
Importantly, wavelengths become shorter over time.

The change in wavelength of M-I waves is a novel manifestation of frequency-based phase mixing.
Since the background values of $E_y$ and $N$ vary with $y$, so too does $f_{MI}$ (see equation \ref{eq:frequency}),
as is apparent from comparing oscillations at $y/y_0=0.52$ and $y/y_0=0.54$ (Figure \ref{fig:periods}).
It follows that M-I waves, formed on the boundary with wavelengths slightly 
larger than the electron inertial length at the base of the magnetosphere, 
phase mix to ever smaller length scales.
Frequency-based phase mixing has been well studied for field line resonances \citep[e.g.][]{1995Mann,2005Rankin,2010RussellWrightAA}
but we apply it for the first time to M-I waves, drawing on the result,
\begin{eqnarray}
 \lambda_y &\sim& \frac{1}{\left(t|df_{MI}/dy|\right)}.\label{eq:phase-mixing}
\end{eqnarray}

We find the wavelength prior to phase mixing (the length scale immediately trailing the broadening front) 
can be estimated by equating the speed of an equivalent discontinuity with the phase speed of the trailing disturbances.
This equality ensures that the broadening front and the undershoot immediately behind it move together as a single structure.
 
In the ideal limit, a discontinuity moves at a speed given by
\begin{eqnarray}
v_d&=&\sqrt{v_{MI}(a)v_{MI}(b)},\label{eq:vd}
\end{eqnarray}
where ``a" (``b") indicates a function or value is evaluated ahead of (behind) the discontinuity
and $v_{MI}$ is given by equation (\ref{eq:v_MI}) \citep{2012RussellWright}.
When electron inertia is included, the broadening front is smoothed from a discontinuity, 
however equation (\ref{eq:vd}) may be used as a first approximation.
Ripples behind the broadening front evolve as M-I waves, which have a phase speed given by \citet{2012RussellWright} as
\begin{eqnarray}
v_p &=& \frac{M_PE_y}{1+(\Sigma_P/\Sigma_A)\sqrt{1+k_y^2\lambda_e^2}}.\label{eq:vp}
\end{eqnarray}

Equating $v_p$ (equation (\ref{eq:vd})) and $v_d$ (equation (\ref{eq:vp})),
and making use of the relationship $E_y=(1+r)E_i$ ,
where $r=(1-\Sigma_P/\Sigma_A)/(1+\Sigma_P/\Sigma_A)$ is the electric field reflection coefficient
and the incident electric field $E_i$ is approximately the same ahead and behind the broadening front,
the wavelength of disturbances produced by broadening is
\begin{eqnarray}
 \lambda_y&=&\frac{2\pi\lambda_e}{\sqrt{(N(a)/N(b))^2-1}},\label{eq:lambda}
\end{eqnarray}
being determined solely by $\lambda_e$ and the ratio of number densities (equivalently Pedersen conductances)
ahead of and behind the broadening front.

Equation (\ref{eq:lambda}) can be tested against the inertial simulation presented in Figure \ref{fig:inertial}.
At time $t/\tau=2.2$, the wavelength behind the broadening front is estimated from perturbed number density
as twice the distance between the local minimum immediately following the front and the following local maximum.
This gives $\lambda_y/\lambda_e = 2.0$.
The number density behind the front is easily identified as $N(b)/N_e=0.312$ (the local minimum).
For this simulation, the value ahead of the front is less precisely determined
because there is some ambiguity about where the transition begins.
We therefore give the range $0.8<N_a/N_e<1.1$, for which equation (\ref{eq:lambda}) returns
$1.86<\lambda_y/\lambda_e<2.66$, in good agreement with $\lambda_y/\lambda_e=2.0$.

This wavelength analysis can also be applied when an initially large-scale M-I wave steepens nonlinearly
to form ripples behind a steep gradient, 
as seen in Figure 2 of \citet{2002LysakSong} and Figure 6 of \citet{2012RussellWright}.
In the latter case, small scales have formed by $t/\tau=4$,
at which time the density ratio across the acting discontinuity is $N(a)/N(b)=1.284$
(the value for $N(a)$ is precisely determined from a local maximum ahead of the main transition).
Therefore, equation (\ref{eq:lambda}) returns $\lambda_y/\lambda_e=7.8$, 
which agrees exactly with the wavelength measured from crossings of $N$ with $N_e$.
This agreement gives further confidence in equation (\ref{eq:lambda}) and demonstrates its wider applicability.
Equation (\ref{eq:lambda}) also explains why $\lambda_y/\lambda_e$ is nearly four times larger
for the wavebreaking simulation of \citet{2012RussellWright} than for the ionospheric depletion scenario studied in the present work:
depletion leads to much greater $N(a)/N(b)$, so $\lambda_y/\lambda_e$ is scaled accordingly.

Thus, production of small scales is a two stage process.
First, strong downward FAC causes a rapid collapse of horizontal scales,
populating the M-I boundary with waves that have a horizontal scale of approximately 
the electron inertial length for the low-altitude magnetosphere,
which drive upgoing inertial Alfv\'en waves.
Thereafter, frequency-based phase mixing of boundary waves continuously reduces horizontal scales further.

\section{Discussion}\label{sec:disc}

The present work shows that small-scale waves are produced by the self-consistent M-I response to strong downward current,
even when amplification due to IFI is excluded.  
This suggests that instability is not responsible for producing these waves, although nonlinear M-I coupling is most definitely at work.
We propose that small scales are actively created by a nonlinear steepening/wavebreaking process that follows ionospheric depletion,
which is also responsible for expanding the ionospheric cavity and associated downward current channel in the direction of the electric field.
In this view, broadening rapidly collapses the shortest horizontal gradient length scale to approximately
the electron inertial length for the low-altitude magnetosphere,
the exact scale determined by the ratio of ionospheric number densities 
(equivalently Pedersen conductances) to either side of the broadening front.
After inertial waves have been formed,
length scales are further diminished by frequency-based phase mixing of M-I boundary waves.
This interpretation contrasts with a previously held view that ionospheric depletion simply creates favorable conditions for IFI,
and leads to different conclusions about resulting wave properties.

In the new interpretation, broadening produces boundary waves with the properties of inertial M-I waves,
which oscillate at a frequency determined by magnetospheric and ionospheric properties.
This frequency provides a central test as to whether this mechanism may 
explain previous observations of small-scale waves in downward current channels.

We shall estimate this frequency under conditions appropriate to the observations of \citet{2004Karlsson}, 
for which wave periods were reported as 20 to 40 seconds.
At the time of observation, the Cluster satellites were passing over the nighttime winter southern hemisphere 
(magnetic footpoints at about 20:00 MLT and $70^\circ{\rm S}$ geomagnetic latitude at about 05:30 UT on 19 May 2002).
Under these conditions, E and F regions are normally well defined, 
so we will assume the sheet ionosphere in our model represents the E-region, 
and properties corresponding to the low altitude magnetosphere will be evaluated at electron density minimum between the E and F regions.
Using typical values,
we take magnetic field strength $B=5.5\times10^{-5}\mbox{ T}$, 
a low-altitude magnetospheric electron density of $n_{mag}=5\times10^8\mbox{ m}^{-3}$
and an average ion mass $m_i=30m_p$ where $m_p$ is the proton mass
(a good approximation for a plasma dominated by $\rm{NO}^+$ followed by $\rm{O}_2^+$).
This gives an electron inertial length of 240 m and an Alfv\'en conductance, $\Sigma_A=0.081\mbox{ mho}$.
Putting $M_P=10^4\mbox{ m}^2{\rm s}^{-1}{\rm V}^{-1}$, $\Sigma_P=0.6\mbox{ mho}$ and $E_y=0.1\mbox{ V}$
(for quantities in the depleted E-region), equation (\ref{eq:frequency}) returns a period of 11 seconds.
This is very close to the observed period of 20 to 40 seconds
and certainly within the uncertainty resulting from plausible deviations from our choice of typical values
(which in any case will vary across the region populated with M-I waves).

We also note that the F-region commonly becomes depleted in association with E-region depletion,
through motion of plasma along the magnetic field and enhanced recombination due to Ohmic heating (as discussed in Section \ref{sec:model}).
Since $f_{MI}\propto n_{mag}$ this effect can easily increase periods by a factor 2 or 3 to agree exactly with those reported by \citet{2004Karlsson}.
In fact, in the real system, we speculate that one may find wave periods becoming longer over several minutes as the F-region depletes.

From the frequency of M-I waves and the perpendicular wavelength at the top of the ionosphere,
one can deduce the properties of upgoing inertial Alfv\'en waves produced by the ionospheric depletion mechanism.
This provides useful tests for identifying observations with this theory.
We start by noting that since M-I waves have a phase speed in the direction of the ionospheric electric field (equation (\ref{eq:vp})),
the phase-speed of the magnetospheric waves should also have a component parallel to the electric field (never anti-parallel).

Wavelengths in the magnetosphere are now considered. 
As an inertial Alfv\'en wave propagates out into the magnetosphere, 
its wavelength perpendicular to the magnetic field may increase with height
due to expansion of the flux tube, and also due to spreading caused by the perpendicular group velocity of the waves (seen in Figure \ref{fig:inertial}).
Spreading due to group velocity can be estimated from the ray angles at the edges of the upgoing wavepacket.
Using the dispersion relation for inertial Alfv\'en waves,
\begin{eqnarray}
 \omega&=&\frac{k_{||}v_A}{\sqrt{1+k_\perp^2\lambda_e^2}},\label{eq:IAW}
\end{eqnarray}
and the group speed expressions $v_{g,\perp}=\partial\omega/\partial k_\perp$, $v_{g,||}=\partial\omega/\partial k_{||}$,
one finds
\begin{eqnarray}
 \frac{v_{g,\perp}}{v_{g,||}}&=&\frac{\omega\lambda_e}{v_A}\left(\frac{k_\perp\lambda_e}{\sqrt{1+k_\perp^2\lambda_e^2}}\right).\label{eq:spread}
\end{eqnarray}
Setting $\omega=2\pi f_{MI}$ under a WKB assumption and putting typical values for $\lambda_e$ and $v_A$,
equation (\ref{eq:spread}) returns values for $v_{g,\perp}/v_{g,||}$ on the order of $10^{-4}$ or less 
(the term in brackets is a factor between zero and one).
Thus, rays are close to field aligned, so that expansion of the magnetic field with height is likely to dominate 
over spreading due to transverse group velocity.

If transverse Alfv\'en speed gradients are present in the magnetosphere
then increases to transverse wavelengths of upgoing inertial Alfv\'en waves
may be partially countered by velocity-based phase mixing, which acts to reduce transverse wavelengths \citep{2004Genot,2008LysakSong}.
These arguments suggest that small-scale waves observed in the magnetosphere
should have transverse wavelengths that map to an ionospheric wavelength similar to or less than the electron inertial length in the low altitude
magnetosphere if they are produced by the depletion mechanism we have identified.
This should be verified in future by more detailed modeling designed to capture changes to the upgoing waves with height,
at which stage damping processes, such as Landau damping, may also be considered.

The models used in this paper have treated the M-I system as having a uniform magnetosphere overlying a sheet ionosphere.
Upgoing Alfv\'en waves were therefore able to travel unhindered to the outer magnetosphere.
In reality, the F-region produces a steep gradient of Alfv\'en speed, which may partially reflect Alfv\'en waves.
Thus, Alfv\'en waves may be partially trapped between the E and F regions, 
undergoing multiple reflections in a volume known as the ionospheric Alfv\'en resonator (IAR) \citep{1991Lysak}.

To see how F-region trapping affects production and evolution of small-scale Alfv\'en waves,
it is useful to compare our findings with the simulations of \citet{2004StreltsovLotko}, e.g. their Figures 3 and 4.
Their study included a non-uniform magnetosphere, F-region, 2D dipole field and 2-fluid effects, 
and its output may be compared to the simulations presented earlier in Sections \ref{sec:waves_without_ifi} and \ref{sec:ideal}.

Many features of the evolution are shared with our models: 
FACs modify E-region plasma density,
downward FAC is unsustainable, depleting the E-region and leading to broadening,
and this, combined with electron inertia, populates a part of the M-I boundary with M-I waves
that drive upgoing inertial Alfv\'en waves that reach into the magnetosphere.
These similarities suggest that our new interpretation,
although founded on comparatively simple M-I models,
does indeed hold when additional physics is included.

The primary difference is that the simulations of \citet{2004StreltsovLotko} 
show small-scale ionospheric disturbances gradually spreading over time,
coming to occupy a substantial part of the downward current channel.
This can be understood as follows:
waves produced by M-I interactions have a component of phase-speed in the direction of the background electric field;
therefore, inertial Alfv\'en waves that are driven by M-I interactions 
have a component of group velocity in the direction opposite to the horizontal electric field.
Consequently, during multiple reflections inside the IAR, 
inertial Alfv\'en waves will slowly spread across the downward FAC, 
modifying the ionosphere and extending the part of the boundary populated with M-I waves.

We also note that IFI is able to act when F-region reflections are present.
Thus, although IFI may not be responsible for the initial creation of small-scale waves,
it could potentially affect their later evolution.
The effects of IFI will depend on the steepness of the F-region Alfv\'en speed gradient producing the trapping,
and also the recombinative decay time in the E-region.
If decay due to E-region effects is more rapid than growth due to IFI,
we speculate that IFI will simply prolong the lifetime of small-scale waves by counteracting their decay.
If IFI is strong enough to overcome decay processes, then it may significantly amplify small-scale waves
that have been produced by the processes described in this paper. 

More sophisticated modeling of the phenomena described in this paper is intended in future work.
For example, our use of a height-integrated ionosphere, 
sufficient to capture the most fundamental aspects of the present problem,
nonetheless obscures issues such as at what altitude the ionospheric density depletions are most extreme
and the height profile of the disturbances that form the ionospheric portion of the M-I waves.
Progressing to height-resolved models that include these features is therefore an important goal for future theoretical work.

Other obvious developments are to remove the 2D assumption made in this paper and to allow nonlinear effects in the magnetosphere.
We speculate that events in such simulations (using an initially 2D setup as in the present paper)
would follow a similar initial evolution to those shown here.
Once small-scale waves form, however, these narrow sheets will be subject to instabilities, 
including shear and tearing mode instabilities \citep{1990Seyler,2011Chaston}.
These may break up the 2D structures into filaments with azimuthal structure, for which M-I wave properties are less clear.

It is also desirable that a detailed comparison to observations be made.
An ideal study would combine measurements of small-scale waves inside a large-scale downward current channel
with simultaneous data from the E and F regions at the magnetic footpoint. 
The tests that would associate these waves with the processes described in this paper are:
close agreement between measured (non Doppler shifted) frequencies and $f_{MI}$ computed from measurements;
a transverse length scale that maps to approximately the electron inertial length at the base of the magnetosphere;
$\vec{v}_{phase}\cdot\vec{E}_\perp>0$; 
and an upward Poynting flux carried by the small-scale waves, 
although care is needed because the total Poynting flux 
(due to small-scale waves and the large-scale driver) is downwards.

To conclude this paper, we comment that
understanding small-scale waves produced by M-I interactions could be significant for a number of disciplines.
One example is prediction of ionospheric densities, which impact on communications, GPS and satellite drag.
Changes in ionospheric number density, caused by FAC, are obvious in our simulations,
but there is an additional aspect that we have not considered:
if small-scale Alfv\'en waves attain sufficient amplitude, then they will exert a ponderomotive force on plasma above the E-region,
moving it along the background magnetic field \citep{1990Boehm,2008StreltsovLotko}.
Our results clarify a mechanism by which Alfv\'en waves performing this role may be produced, 
and constrain their frequency and wavelengths.

The processes described here may also explain features of auroral dynamics.
Further work is needed to model the auroral signatures of these waves, but the evolving current patterns
are a guide to what may be expected.
If one associates strong upward currents with optical auroral emission,
one can imagine a preexisting auroral arc brightening as the large-scale current supplying it with electrons intensifies.
Were current density in the corresponding downward FAC to increase beyond the critical threshold, 
the dark region would expand into the preexisting arc,
while new narrow auroral arcs, associated with the processes we have described, formed at its edge.
As we have seen, the narrow arcs would move in the direction of the large-scale electric field (with the phase motion of the waves)
but could spread in the opposite direction (due to reflections of inertial Alfv\'en waves in the IAR).
Such a scenario is reminiscent of observations by \citet{2008Semeter} and
would be an interesting topic for future investigation.

%%%%%%%%%%%%%%%%%%%%%%%%%%%%%%%%%%%%%%%%%%%%%%%%%%%%%%%%%%%%%%%%%%%%%%%%%%%%%%%%%%%%%%%%%%%%%%%%%%%%%%%%%

\begin{acknowledgments}
The authors are grateful to the International Space Science Institute (Switzerland) for funding the program that inspired this work.  
AJBR is grateful to the Royal Commission for the Exhibition of 1851 for present support and acknowledges an STFC studentship that funded part of this work.
\end{acknowledgments}

\bibliographystyle{agu08}
%\bibliography{AWorigin}

\begin{thebibliography}{39}
\providecommand{\natexlab}[1]{#1}
\expandafter\ifx\csname urlstyle\endcsname\relax
  \providecommand{\doi}[1]{doi:\discretionary{}{}{}#1}\else
  \providecommand{\doi}{doi:\discretionary{}{}{}\begingroup
  \urlstyle{rm}\Url}\fi

\bibitem[{\textit{{Atkinson}}(1970)}]{1970Atkinson}
{Atkinson}, G. (1970), {Auroral arcs: Result of the interaction of a dynamic
  magnetosphere with the ionosphere.}, \textit{J. Geophys. Res.}, \textit{75},
  4746--4755, \doi{10.1029/JA075i025p04746}.

\bibitem[{\textit{{Blixt} and {Brekke}}(1996)}]{1996BlixtBrekke}
{Blixt}, E.~M., and A.~{Brekke} (1996), {A model of currents and electric
  fields in a discrete auroral arc}, \textit{Geophys. Res. Lett.}, \textit{23},
  2553--2556, \doi{10.1029/96GL02378}.

\bibitem[{\textit{{Boehm} et~al.}(1990)\textit{{Boehm}, {Carlson}, {McFadden},
  {Clemmons}, and {Mozer}}}]{1990Boehm}
{Boehm}, M.~H., C.~W. {Carlson}, J.~P. {McFadden}, J.~H. {Clemmons}, and F.~S.
  {Mozer} (1990), {High-resolution sounding rocket observations of
  large-amplitude Alfven waves}, \textit{Journal of Geophysical Research (Space
  Physics)}, \textit{95}, 12,157--12,171, \doi{10.1029/JA095iA08p12157}.

\bibitem[{\textit{Brekke}(1997)}]{1997Brekke}
Brekke, A. (1997), \textit{Physics of the Upper Polar Atmosphere}, Wiley-Praxis
  Series in Atmospheric Physics, John Wiley \& Sons.

\bibitem[{\textit{{Chaston} et~al.}(2002)\textit{{Chaston}, {Bonnell},
  {Peticolas}, {Carlson}, {McFadden}, and {Ergun}}}]{2002Chaston}
{Chaston}, C.~C., J.~W. {Bonnell}, L.~M. {Peticolas}, C.~W. {Carlson}, J.~P.
  {McFadden}, and R.~E. {Ergun} (2002), {Driven Alfven waves and electron
  acceleration: A FAST case study}, \textit{Geophys. Res. Lett.},
  \textit{29}(11), 1535, \doi{10.1029/2001GL013842}.

\bibitem[{\textit{{Chaston} et~al.}(2011)\textit{{Chaston}, {Seki}, {Sakanoi},
  {Asamura}, {Hirahara}, and {Carlson}}}]{2011Chaston}
{Chaston}, C.~C., K.~{Seki}, T.~{Sakanoi}, K.~{Asamura}, M.~{Hirahara}, and
  C.~W. {Carlson} (2011), {Cross-scale coupling in the auroral acceleration
  region}, \textit{\grl}, \textit{38}, L20101, \doi{10.1029/2011GL049185}.

\bibitem[{\textit{{Cran-McGreehin} et~al.}(2007)\textit{{Cran-McGreehin},
  {Wright}, and {Hood}}}]{2007CranMcGreehin}
{Cran-McGreehin}, A.~P., A.~N. {Wright}, and A.~W. {Hood} (2007), {Ionospheric
  depletion in auroral downward currents}, \textit{Journal of Geophysical
  Research (Space Physics)}, \textit{112}(A11), 10,309,
  \doi{10.1029/2007JA012350}.

\bibitem[{\textit{{Doe} et~al.}(1995)\textit{{Doe}, {Vickrey}, and
  {Mendillo}}}]{1995Doe}
{Doe}, R.~A., J.~F. {Vickrey}, and M.~{Mendillo} (1995), {Electrodynamic model
  for the formation of auroral ionospheric cavities}, \textit{J. Geophys.
  Res.}, \textit{100}, 9683--9696, \doi{10.1029/95JA00001}.

\bibitem[{\textit{{Dreher}}(1997)}]{1997Dreher}
{Dreher}, J. (1997), {On the self-consistent description of dynamic
  magnetosphere-ionosphere coupling phenomena with resolved ionosphere},
  \textit{J. Geophys. Res.}, \textit{102}, 85--94, \doi{10.1029/96JA02800}.

\bibitem[{\textit{{G{\'e}not} et~al.}(2004)\textit{{G{\'e}not}, {Louarn}, and
  {Mottez}}}]{2004Genot}
{G{\'e}not}, V., P.~{Louarn}, and F.~{Mottez} (2004), {Alfv{\'e}n wave
  interaction with inhomogeneous plasmas: acceleration and energy cascade
  towards small-scales}, \textit{Annales Geophysicae}, \textit{22}, 2081--2096,
  \doi{10.5194/angeo-22-2081-2004}.

\bibitem[{\textit{{Holzer} and {Sato}}(1973)}]{1973HolzerSato}
{Holzer}, T.~E., and T.~{Sato} (1973), {Quiet auroral arcs and electrodynamic
  coupling between the ionosphere and the magnetosphere. 2.}, \textit{J.
  Geophys. Res.}, \textit{78}, 7330--7339, \doi{10.1029/JA078i031p07330}.

\bibitem[{\textit{{Johansson} et~al.}(2004)\textit{{Johansson}, {Figueiredo},
  {Karlsson}, {Marklund}, {Fazakerley}, {Buchert}, {Lindqvist}, and
  {Nilsson}}}]{2004Johansson}
{Johansson}, T., S.~{Figueiredo}, T.~{Karlsson}, G.~{Marklund},
  A.~{Fazakerley}, S.~{Buchert}, P.~{Lindqvist}, and H.~{Nilsson} (2004),
  {Intense high-altitude auroral electric fields - temporal and spatial
  characteristics}, \textit{Annales Geophysicae}, \textit{22}, 2485--2495.

\bibitem[{\textit{{Karlsson} and {Marklund}}(1998)}]{1998KarlssonMarklund}
{Karlsson}, T., and G.~T. {Marklund} (1998), {Simulations of effects of
  small-scale auroral current closure in the return current region},
  \textit{{Phys. Space Plasmas}}, \textit{15}, 401.

\bibitem[{\textit{{Karlsson} et~al.}(2004)\textit{{Karlsson}, {Marklund},
  {Figueiredo}, {Johansson}, and {Buchert}}}]{2004Karlsson}
{Karlsson}, T., G.~{Marklund}, S.~{Figueiredo}, T.~{Johansson}, and
  S.~{Buchert} (2004), {Separating spatial and temporal variations in auroral
  electric and magnetic fields by Cluster multipoint measurements},
  \textit{Annales Geophysicae}, \textit{22}, 2463--2472.

\bibitem[{\textit{{Keiling} et~al.}(2005)\textit{{Keiling}, {Parks}, {Wygant},
  {Dombeck}, {Mozer}, {Russell}, {Streltsov}, and {Lotko}}}]{2005Keiling}
{Keiling}, A., G.~K. {Parks}, J.~R. {Wygant}, J.~{Dombeck}, F.~S. {Mozer},
  C.~T. {Russell}, A.~V. {Streltsov}, and W.~{Lotko} (2005), {Some properties
  of Alfv{\'e}n waves: Observations in the tail lobes and the plasma sheet
  boundary layer}, \textit{Journal of Geophysical Research (Space Physics)},
  \textit{110}(A9), 10, \doi{10.1029/2004JA010907}.

\bibitem[{\textit{{Lysak}}(1991)}]{1991Lysak}
{Lysak}, R.~L. (1991), {Feedback instability of the ionospheric resonant
  cavity}, \textit{J. Geophys. Res.}, \textit{96}, 1553--1568,
  \doi{10.1029/90JA02154}.

\bibitem[{\textit{{Lysak} and {Song}}(2002)}]{2002LysakSong}
{Lysak}, R.~L., and Y.~{Song} (2002), {Energetics of the ionospheric feedback
  interaction}, \textit{Journal of Geophysical Research (Space Physics)},
  \textit{107}, 1160, \doi{10.1029/2001JA000308}.

\bibitem[{\textit{{Lysak} and {Song}}(2008)}]{2008LysakSong}
{Lysak}, R.~L., and Y.~{Song} (2008), {Propagation of kinetic Alfv{\'e}n waves
  in the ionospheric Alfv{\'e}n resonator in the presence of density cavities},
  \textit{Geophys. Res. Lett.}, \textit{35}, L20101,
  \doi{10.1029/2008GL035728}.

\bibitem[{\textit{{Mann} et~al.}(1995)\textit{{Mann}, {Wright}, and
  {Cally}}}]{1995Mann}
{Mann}, I.~R., A.~N. {Wright}, and P.~S. {Cally} (1995), {Coupling of
  magnetospheric cavity modes to field line resonances: A study of resonance
  widths}, \textit{J. Geophys. Res.}, \textit{100}, 19,441--19,456,
  \doi{10.1029/95JA00820}.

\bibitem[{\textit{{Rankin} et~al.}(2005)\textit{{Rankin}, {Kabin}, {Lu},
  {Mann}, {Marchand}, {Rae}, {Tikhonchuk}, and {Donovan}}}]{2005Rankin}
{Rankin}, R., K.~{Kabin}, J.~Y. {Lu}, I.~R. {Mann}, R.~{Marchand}, I.~J. {Rae},
  V.~T. {Tikhonchuk}, and E.~F. {Donovan} (2005), {Magnetospheric field-line
  resonances: Ground-based observations and modeling}, \textit{Journal of
  Geophysical Research (Space Physics)}, \textit{110}(A9), A10S09,
  \doi{10.1029/2004JA010919}.

\bibitem[{\textit{{Russell}}(2010)}]{2010Russell}
{Russell}, A.~J.~B. (2010), {Coupling of the Solar Wind, Magnetosphere and
  Ionosphere by MHD Waves}, Ph.D. thesis, Univ. of St Andrews, St Andrews,
  Fife, UK.

\bibitem[{\textit{{Russell} and {Wright}}(2010)}]{2010RussellWrightAA}
{Russell}, A.~J.~B., and A.~N. {Wright} (2010), {Resonant absorption with 2D
  variation of field line eigenfrequencies}, \textit{Astron. Astrophys.},
  \textit{511}, A17, \doi{10.1051/0004-6361/200912669}.

\bibitem[{\textit{{Russell} and {Wright}}(2012)}]{2012RussellWright}
{Russell}, A.~J.~B., and A.~N. {Wright} (2012), {Magnetosphere-ionosphere
  waves}, \textit{Journal of Geophysical Research (Space Physics)},
  \textit{117}(A16), A01202, \doi{10.1029/2011JA016950}.

\bibitem[{\textit{{Russell} et~al.}(2010)\textit{{Russell}, {Wright}, and
  {Hood}}}]{2010RussellWright}
{Russell}, A.~J.~B., A.~N. {Wright}, and A.~W. {Hood} (2010), {Self-consistent
  ionospheric plasma density modifications by field-aligned currents: Steady
  state solutions}, \textit{Journal of Geophysical Research (Space Physics)},
  \textit{115}(A14), 4216, \doi{10.1029/2009JA014836}.

\bibitem[{\textit{{Saint Maurice} and {Torr}}(1978)}]{1978StMauriceTorr}
{Saint Maurice}, J.-P., and D.~G. {Torr} (1978), {Nonthermal rate coefficients
  in the ionosphere - The reactions of O/+/ with N2, O2, and NO}, \textit{J.
  Geophys. Res.}, \textit{83}, 969--977, \doi{10.1029/JA083iA03p00969}.

\bibitem[{\textit{{Sato}}(1978)}]{1978Sato}
{Sato}, T. (1978), {A theory of quiet auroral arcs}, \textit{J. Geophys. Res.},
  \textit{83}, 1042--1048, \doi{10.1029/JA083iA03p01042}.

\bibitem[{\textit{{Semeter} et~al.}(2008)\textit{{Semeter}, {Zettergren},
  {Diaz}, and {Mende}}}]{2008Semeter}
{Semeter}, J., M.~{Zettergren}, M.~{Diaz}, and S.~{Mende} (2008), {Wave
  dispersion and the discrete aurora: New constraints derived from high-speed
  imagery}, \textit{Journal of Geophysical Research (Space Physics)},
  \textit{113}(A12), A12,208, \doi{10.1029/2008JA013122}.

\bibitem[{\textit{{Seyler}}(1990)}]{1990Seyler}
{Seyler}, C.~E. (1990), {A mathematical model of the structure and evolution of
  small-scale discrete auroral arcs}, \textit{J. Geophys. Res.}, \textit{95},
  17,199--17,215, \doi{10.1029/JA095iA10p17199}.

\bibitem[{\textit{{Stasiewicz} et~al.}(2000)}]{2000Stasiewicz}
{Stasiewicz}, K., et~al. (2000), {Small Scale Alfv{\'e}nic Structure in the
  Aurora}, \textit{Space Science Reviews}, \textit{92}, 423--533.

\bibitem[{\textit{{Streltsov} and {Karlsson}}(2008)}]{2008StreltsovKarlsson}
{Streltsov}, A.~V., and T.~{Karlsson} (2008), {Small-scale, localized
  electromagnetic waves observed by Cluster: Result of magnetosphere-ionosphere
  interactions}, \textit{Geophys. Res. Lett.}, \textit{35}, 22,107,
  \doi{10.1029/2008GL035956}.

\bibitem[{\textit{{Streltsov} and {Lotko}}(2004)}]{2004StreltsovLotko}
{Streltsov}, A.~V., and W.~{Lotko} (2004), {Multiscale electrodynamics of the
  ionosphere-magnetosphere system}, \textit{Journal of Geophysical Research
  (Space Physics)}, \textit{109}(A18), 9214, \doi{10.1029/2004JA010457}.

\bibitem[{\textit{{Streltsov} and {Lotko}}(2008)}]{2008StreltsovLotko}
{Streltsov}, A.~V., and W.~{Lotko} (2008), {Coupling between density
  structures, electromagnetic waves and ionospheric feedback in the auroral
  zone}, \textit{Journal of Geophysical Research (Space Physics)},
  \textit{113}(A12), A05212, \doi{10.1029/2007JA012594}.

\bibitem[{\textit{{Streltsov} et~al.}(2010)\textit{{Streltsov}, {Pedersen},
  {Mishin}, and {Snyder}}}]{Streltsov_substorm}
{Streltsov}, A.~V., T.~R. {Pedersen}, E.~V. {Mishin}, and A.~L. {Snyder}
  (2010), {Ionospheric feedback instability and substorm development},
  \textit{Journal of Geophysical Research (Space Physics)}, \textit{115}(A14),
  7205, \doi{10.1029/2009JA014961}.

\bibitem[{\textit{{Sydorenko} et~al.}(2008)\textit{{Sydorenko}, {Rankin}, and
  {Kabin}}}]{2008Sydorenko}
{Sydorenko}, D., R.~{Rankin}, and K.~{Kabin} (2008), {Nonlinear effects in the
  ionospheric Alfv{\'e}n resonator}, \textit{Journal of Geophysical Research
  (Space Physics)}, \textit{113}(A12), A10206, \doi{10.1029/2008JA013579}.

\bibitem[{\textit{{Trakhtengerts} and
  {Feldstein}}(1984)}]{1984TrakhtengertsFeldsein}
{Trakhtengerts}, V.~I., and A.~I. {Feldstein} (1984), {Quiet auroral arcs -
  Ionosphere effect of magnetospheric convection stratification},
  \textit{Planet. Space Sci.}, \textit{32}, 127--134,
  \doi{10.1016/0032-0633(84)90147-8}.

\bibitem[{\textit{{Wright} et~al.}(2008)\textit{{Wright}, {Owen}, {Chaston},
  and {Dunlop}}}]{2008Wright}
{Wright}, A.~N., C.~J. {Owen}, C.~C. {Chaston}, and M.~W. {Dunlop} (2008),
  {Downward current electron beam observed by Cluster and FAST},
  \textit{Journal of Geophysical Research (Space Physics)}, \textit{113}(A12),
  6202, \doi{10.1029/2007JA012643}.

\bibitem[{\textit{{Zettergren} and {Semeter}}(2012)}]{2012ZettergrenSemeter}
{Zettergren}, M., and J.~{Semeter} (2012), {Ionospheric plasma transport and
  loss in auroral downward current regions}, \textit{Journal of Geophysical
  Research (Space Physics)}, \textit{117}(A16), A06306,
  \doi{10.1029/2012JA017637}.

\bibitem[{\textit{{Zettergren} et~al.}(2010)\textit{{Zettergren}, {Semeter},
  {Burnett}, {Oliver}, {Heinselman}, {Blelly}, and {Diaz}}}]{2010Zettergren}
{Zettergren}, M., J.~{Semeter}, B.~{Burnett}, W.~{Oliver}, C.~{Heinselman},
  P.-L. {Blelly}, and M.~{Diaz} (2010), {Dynamic variability in F-region
  ionospheric composition at auroral arc boundaries}, \textit{Annales
  Geophysicae}, \textit{28}, 651--664, \doi{10.5194/angeo-28-651-2010}.

\bibitem[{\textit{{Zhu} et~al.}(2001)\textit{{Zhu}, {Otto}, {Lummerzheim},
  {Rees}, and {Lanchester}}}]{2001Zhu}
{Zhu}, H., A.~{Otto}, D.~{Lummerzheim}, M.~H. {Rees}, and B.~S. {Lanchester}
  (2001), {Ionosphere-magnetosphere simulation of small-scale structure and
  dynamics}, \textit{J. Geophys. Res.}, \textit{106}, 1795--1806,
  \doi{10.1029/1999JA000291}.

\end{thebibliography}

\end{article}
\end{document}